\documentclass{article}
\def\s{\sigma}

\def\be{\begin{equation}}
\def\ee{\end{equation}}
\def\arr{\begin{array}{rll}}
\def\ea{\end{array}}
\def\bea{\begin{eqnarray}}
\def\eea{\end{eqnarray}}

\def\N2{$N{=}2$}

\def\>{\rangle}
\def\<{\langle}
\def\+{\dagger}
\def\={\ =\ }

\begin{document}
\begin{center}
{\Large\bf  Superintegrable models related to near horizon extremal }\\
\vskip 0.3cm
{\Large\bf    Myers--Perry black hole in arbitrary dimension }\\

\vspace{0.5 cm} {\large
 Anton Galajinsky$^1$, Armen Nersessian$^2$, Armen Saghatelian$^2$
}
\end{center}
{\sl
$^1$Tomsk Polytechnic University, Lenin Ave. 30, 634050 Tomsk, Russia\\
$^2$ Yerevan State University,
1 Alex Manoogian St., Yerevan, 0025, Armenia}

\begin{abstract} \noindent
We provide a systematic account of integrability of the spherical mechanics associated with
the near horizon extremal  Myers--Perry black hole in arbitrary dimension for the special case that all rotation parameters are equal.
The integrability is established both in the original coordinates and in action--angle variables.
It is demonstrated that the spherical mechanics associated with the black hole in $d=2n+1$ is maximally superintegrable, while its counterpart related to the black hole in $d=2n$
lacks for only one integral of motion to be maximally superintegrable.
\end{abstract}
%

\renewcommand{\thefootnote}{\arabic{footnote}}
\setcounter{footnote}0
\section{Introduction}

Models of conformal mechanics associated with near horizon geometry of extremal black holes in
diverse dimensions are being extensively studied for more than a decade (see e.g. \cite{kallosh}--\cite{gal}
 and references therein).
There are several reasons to be concerned
about such systems. On the one hand, they provide a useful means of studying geometry of vacuum solutions
of the Einstein equations.
For example, the geodesic equations for a massive particle in the Kerr space--time admit a quadratic
first integral \cite{car}, which can be linked to
the second rank Killing tensor \cite{wp}. The use of conformal mechanics enables one to establish its reducibility in the near horizon limit
\cite{g2,go}.
On the other hand, it is expected that some variant of conformal mechanics will turn out to be useful within the
context of the ${AdS}_2/{CFT}_1$--duality and/or the Kerr/CFT--correspondence.

A remarkable property of the near horizon extremal black hole in arbitrary dimension is that its isometry group involves the conformal factor $SO(2,1)$ (see e.g. \cite{bh}).
Because Killing vectors are linked to first integrals of the geodesic equations, a massive relativistic particle propagating on such a background inherits the conformal invariance
and belongs to the class of conformal mechanics models.
A salient feature of this system is that, by applying a suitable canonical transformation, the radial canonical pair can be separated from angular variables and the model
can be put in the conventional conformal mechanics form \cite{bgik,g1,gn}. At the off--shell Lagrangian level the radial motion of the particle was related to $d=1$
conformal mechanics in an earlier paper \cite{ikn}. Because the variables are separated, the angular sector
can be studied in its own right. In particular, the Casimir element of the conformal algebra $so(2,1)$ realized
in the original relativistic particle determines the Hamiltonian of a reduces mechanics \cite{ar}. In what follows
we call it the spherical mechanics. It is important to notice that if the isometry group $G$ of a near horizon
black hole configuration has the direct product structure $SO(2,1) \times H$, where $H$ is a subgroup of $G$,
then the spherical mechanics inherits the invariance under the action of $H$.

Although the spherical mechanics related to nonrelativistic conformal many--body models has been extensively investigated
in a series of works \cite{ar}-\cite{hlnsy}, systems originating from near horizon extremal black hole
geometries remain almost completely unexplored. Yet, it is true that they may provide new nontrivial
examples of superintegrable interacting  models. In a recent work \cite{gal}, the Hamiltonian
of a spherical mechanics associated with the near horizon geometry of the extremal Myers--Perry black hole
in arbitrary dimension has been constructed for the special case that all rotation parameters are equal.
This configuration is maximally symmetric and
possesses $SO(2,1)\times U(n)$  isometry groups for $d=2n+1$ and $SO(2,1)\times U(n-1)$ for $d=2n$, respectively.
While integrability of such a spherical mechanics has been announced in \cite{gal},
no explicit proof has been given. The purpose of this work is to provide a carefully argued and
systematic account of the integrability and, furthermore, the superintegrability\footnote{Recall that a Hamiltonian system with $2n$
phase space degrees of freedom is called Liouville integrable if it admits $n$ functionally
independent integrals of motion in involution. If there are more than $n$ such integrals,
the model is called superintegrable. A maximal number of functionally independent integrals
of motion is $2n-1$. Systems possessing $2n-1$ first integrals are called maximally superintegrable. }
of this model. The analysis is facilitated in spherical coordinates, in which complete separation of variables occurs.
The model related to the extremal rotating black hole in $d=2n+1$ dimensions turns out to be maximally superintegrable,
while its counterpart associated with the black hole in $d=2n$ dimensions lacks for only one first
integral to be maximally superintegrable. The former system is shown to contain the latter as a subsystem, which correlates with the fact that
the background metrics in $d=2n+1$ and $d=2(n+1)$ have the same isometry group. We also construct the action-angle variables, which make
semiclassical quantization of the models immediately feasible.

The work is organized as follows. In Sect. 2 we consider the near horizon metrics describing the extremal Myers--Perry black hole in arbitrary dimension
for the special case that all rotation parameters are equal with a special emphasis on their symmetries. In particular, we demonstrate that for $d=2n+1$ the isometry group is $SO(2,1)\times U(n)$
and for $d=2n$ it is $SO(2,1)\times U(n-1)$. In Sect. 3 the basic idea behind the spherical mechanics is reviewed and a canonical transformation, which splits
the radial canonical pair from the rest and brings the model of a massive relativistic particle moving near the horizon of an extremal black hole to the conventional conformal mechanics form, is given.
Sect. 4 is devoted to a systematic account of a maximal superintegrability of the  spherical
mechanics associated with the near horizon extremal rotating black hole in $d=2n+1$ dimensions. A model related to a similar black hole in $d=2n$ is analyzed in Sect. 5 and is shown to
lack for only one constant of the motion to be maximally superintegrable. We summarize our results and discuss possible further developments in the concluding Sect. 6.

\section{Near horizon metrics and their symmetries }
\subsection{ $d=2n+1$ }

A vacuum solution of the Einstein equations describing the Myers--Perry black hole in $d=2n+1$ dimensions for the special case that all $n$ rotation parameters are equal reads \cite{mp}
\bea\label{ed2}
&&
ds^2=\frac{\Delta}{U} {\left(dt -a \sum_{i=1}^{n} \mu_i^2 d \phi_i\right)}^2-\frac{U}{\Delta} dr^2
-\frac{1}{r^2} \sum_{i=1}^{n} \mu_i^2 {\left(a dt-(r^2+a^2) d\phi_i\right)}^2
\nonumber\\[2pt]
&& \qquad-(r^2+a^2) \sum_{i=1}^{n} d \mu_i^2+\frac{a^2 (r^2+a^2)}{r^2} \sum_{i<j}^{n} \mu_i^2 \mu_j^2 {\left(d\phi_i-d\phi_j \right)}^2,
\\[2pt]
&&
\Delta=\frac{{(r^2+a^2)}^n}{r^2}-2M, \quad U={(r^2+a^2)}^{n-1}, \quad \mu_n^2=1-\sum_{i=1}^{n-1} \mu_i^2,
\nonumber
\eea
where $M$ stands for the mass and $a$ is the rotation parameter. In what follows we focus on the extremal solution, for which
\be
M=\frac{n^n {r_0}^{2n-2}}{2}, \qquad a^2=(n-1) r_0^2.
\ee
These conditions follow from the requirement that $\Delta(r)$ has a double zero at the horizon radius $r=r_0$.

The isometry group of (\ref{ed2}) is $U(1)\times U(n)$. The first factor corresponds to time translations, while the second factor describes the enhanced symmetry $U(1)^n \to U(n)$, which occurs if all
rotation parameters of the black hole are set equal.
In order to make $U(n)$ explicit, one parametrizes $n$ spatial two--planes, in which the black hole may rotate, by the coordinates (see, e.g., Ref. \cite{vsp})
\be\label{planes}
x_i=r \mu_i \cos{\phi_i}, \qquad y_i=r \mu_i \sin{\phi_i},
\ee
where $i=1,\dots,n$, and constructs the vector fields
\be\label{vf}
\xi_{ij}=x_i \frac{\partial}{\partial x_j}-x_j \frac{\partial}{\partial x_i}+y_i \frac{\partial}{\partial y_j}-y_j \frac{\partial}{\partial y_i}, \qquad
\rho_{ij}=x_i \frac{\partial}{\partial y_j}-y_j \frac{\partial}{\partial x_i}+x_j \frac{\partial}{\partial y_i}-y_i \frac{\partial}{\partial x_j}.
\ee
These are antisymmetric and symmetric in their indices, respectively, and obey the structure relations of $u(n)$\footnote{The conventional structure relations of $u(n)$ are derived form (\ref{str}) by considering another basis $E_{ab}=\frac 12 (\xi_{ab}+i \rho_{ab})$, the Casimir elements of $u(n)$ being $C_1=E_{i_1 i_1}$, $C_2=E_{i_1 i_2} E_{i_2 i_1}$, \dots, $C_n=E_{i_1 i_2} E_{i_2 i_3} \dots E_{i_n i_1}$.}
\bea\label{str}
&&
[\xi_{ij},\xi_{rs}]=\delta_{jr} \xi_{is} +\delta_{is} \xi_{jr} -\delta_{ir} \xi_{js}-\delta_{js} \xi_{ir}, \qquad
[\rho_{ij},\rho_{rs}]=-\delta_{jr} \xi_{is} -\delta_{ir} \xi_{js} -\delta_{is} \xi_{jr}-\delta_{js} \xi_{ir},
\nonumber\\[2pt]
&&
[\xi_{ij},\rho_{rs}]=\delta_{jr} \rho_{is} +\delta_{js} \rho_{ir} -\delta_{ir} \rho_{js}-\delta_{is} \rho_{jr}.
\eea
It is straightforward to verify that (\ref{vf}) are the Killing vectors of the original black hole metric.
Another way to reveal the $U(n)$--symmetry  is to introduce the complex coordinates
\be
z_j=r \mu_j e^{i \phi_j}
\ee
and rewrite the metric in terms of them. In the complex notation the unitary symmetry is manifest.

In order to construct the near horizon metric, one redefines the coordinates \cite{gal}
\be
r \quad \rightarrow \quad r_0 + \epsilon r_0 r, \qquad t \quad \rightarrow \quad \frac{n r_0 t}{2(n-1)\epsilon}, \qquad \phi_i \quad \rightarrow \quad \phi_i+\frac{r_0 t}{2 a \epsilon}
\ee
and then sends $\epsilon$ to zero. This yields
\bea\label{Nhm1}
&&
ds^2=r^2 dt^2-\frac{dr^2}{r^2}-2n(n-1)\sum_{i=1}^{n} d\mu_i^2-2 \sum_{i=1}^{n} \mu_i^2 {(r dt+n \sqrt{n-1} d \phi_i)}^2+
\nonumber\\[2pt]
&&
\qquad \quad +2n{(n-1)}^2 \sum_{i<j}^{n} \mu_i^2 \mu_j^2 {(d \phi_i-d\phi_j)}^2, \qquad \mu_n^2=1-\sum_{i=1}^{n-1} \mu_i^2.
\eea
It is straightforward to verify that (\ref{Nhm1}) is a vacuum solution of the Einstein equations.
The near horizon metric has a larger symmetry. In addition to $U(1)\times U(n)$ transformations considered above, the isometry group of (\ref{Nhm1}) includes the
dilatation
\be\label{tp1}
t'=t+\lambda t, \qquad r'=r-\lambda r,
\ee
and the special conformal transformation
\be\label{sp}
t'=t+(t^2+\frac{1}{r^2}) \s, \qquad r'=r-2 tr\s, \qquad
{\phi'}_i=\phi_i-\frac{2}{r n \sqrt{n-1} } \s,
\ee
which all together form $SO(2,1)\times U(n)$, the first factor being the conformal group in one dimension.

\subsection{$d=2n$}

A vacuum solution of the Einstein equations describing the Myers--Perry black hole in $d=2n$ dimensions for the special case that all $n-1$ rotation parameters are equal, reads \cite{mp}
\bea\label{ed1}
&&
ds^2=\frac{\Delta}{U} {\left(dt -a \sum_{i=1}^{n-1} \mu_i^2 d \phi_i\right)}^2-\frac{U}{\Delta} dr^2
-\frac{{(r^2+a^2)}^{n-2}}{r U} \sum_{i=1}^{n-1} \mu_i^2 {\left(a dt-(r^2+a^2) d\phi_i\right)}^2
\nonumber\\[2pt]
&& \qquad-(r^2+a^2) \sum_{i=1}^{n-1} d\mu_i^2-r^2 d \mu_n^2 +\frac{a^2 {(r^2+a^2)}^{n-1}}{r U} \sum_{i<j}^{n-1} \mu_i^2 \mu_j^2 {\left(d\phi_i-d\phi_j \right)}^2,
\\[2pt]
&&
\Delta=\frac{1}{r} {(r^2+a^2)}^{n-1}-2M, \quad U=\frac{1}{r} {(r^2+a^2)}^{n-2} (r^2+a^2 \mu_n^2), \quad \mu_n^2=1-\sum_{i=1}^{n-1} \mu_i^2,
\nonumber
\eea
where $M$ is the mass and $a$ is the rotation parameter. As compared to the previous case, the number of the azimuthal coordinates is decreased by one.
For the extremal solution $\Delta$ has a double zero at the horizon radius $r=r_0$. In particular, from $\Delta(r_0)=0$ and $\Delta'(r_0)=0$ one finds
\be
M=\frac{r_0^{2n-3} {[2(n-1)]}^{n-1}}{2}, \qquad a^2=(2n-3) r_0^2.
\ee

The isometry group of (\ref{ed1}) includes time translations and the enhanced rotational symmetry $U(1)^{n-1} \to U(n-1)$, which is a consequence of setting
all the rotation parameters equal. The unitary symmetry is manifest in the complex coordinates
\be
z_j=\mu_j e^{i \phi_j}=x_j+i y_j.
\ee
The corresponding Killing vector fields are realized as in Eq. (\ref{vf}) with $x_i$ and $y_i$ taken from the previous line.

In order to implement the near horizon limit, one redefines the coordinates
\be
r \quad \rightarrow \quad r_0 + \epsilon r_0 r, \qquad t \quad \rightarrow \quad \frac{2(n-1) r_0 t}{(2n-3)\epsilon}, \qquad \phi_i \quad \rightarrow \quad \phi_i+ \frac{r_0 t}{a \epsilon},
\ee
and then
sends $\epsilon$ to zero, which yields \cite{gal}
\bea\label{Nhm}
&&
ds^2=\rho_0^2 \left(r^2 dt^2-\frac{dr^2}{r^2} \right)-2(n-1)\sum_{i=1}^{n-1} d\mu_i^2-d\mu_n^2+\frac{2 (n-1)}{\rho_0^2} \sum_{i<j}^{n-1} \mu_i^2 \mu_j^2 {(d \phi_i-d\phi_j)}^2-
\nonumber\\[2pt]
&&
\qquad \quad -\frac{4}{{(2n-3)}^2 \rho_0^2} \sum_{i=1}^{n-1} \mu_i^2 {(r dt+(n-1) \sqrt{2n-3}d \phi_i)}^2,
\\[2pt]
&&
\rho_0^2=\frac{1+(2n-3) \mu_n^2}{2n-3}, \qquad \mu_n^2=1-\sum_{i=1}^{n-1} \mu_i^2.
\nonumber
\eea
It is straightforward to verify that (\ref{Nhm}) is a vacuum solution of the Einstein equations.
Like in $d=2n+1$, the near horizon metric exhibits additional conformal symmetry, which
is realized as in Eqs. (\ref{tp1}) and (\ref{sp}) with the obvious alteration of the special conformal transformation
\be
{\phi'}_i=\phi_i-\frac{2}{r (n-1) \sqrt{2n-3} } \s
\ee
acting on the azimuthal angular variables. Thus, for $d=2n$ the near horizon symmetry is $SO(2,1)\times U(n-1)$.

\section{Spherical mechanics}

Spherical mechanics has been introduced in \cite{ar} as a specific sector of a generic Hamiltonian mechanics invariant under the action of $SO(2,1)$ group. In general, conformal mechanics is described by the triple $H$, $D=t H+D_0$, $K=t^2 H+2 t D_0+K_0$, where $D_0={D|}_{t=0}$, $K_0={K|}_{t=0}$ and $t$ is the temporal coordinate,  obeying the structure relations of  $so(2,1)$ algebra under the Poisson bracket
\be
\label{confalg}
\{H,D\}=H, \quad \{H,K \}=2D, \quad \{D,K \} =K.
\ee
$H$ is treated as the Hamiltonian, while $D$ and $K$ are conserved charges corresponding to the dilatations and the special conformal transformations, respectively.
Note that $H$, $D_0$ and $K_0$ obey the structure relations of $so(2,1)$  as well. The latter fact allows one to separate the radial canonical pair from the rest by introducing the new radial coordinate \cite{ar}
\be\label{Rtr}
R=\sqrt{2 K_0}, \qquad p_R=-\frac{2 D_0}{\sqrt{2 K_0}} \quad \Rightarrow \quad \{R,p_R\}=1
\ee
such that
\be\label{Hrtr}
H=\frac 12 p_R^2+\frac{2\mathcal{I}}{R^2},
\ee
where $\mathcal{I}$ is the Casimir element of $so(2,1)$
\be\label{i}
\mathcal{I}=H K-D^2=H K_0-D_0^2.
\ee
In general, $\mathcal{I}$ is at most quadratic in momenta canonically conjugate to the remaining angular variables. For this reason it can be viewed as the Hamiltonian of a reduced mechanics, called in \cite{ar}
the spherical mechanics.

As was demonstrated above, the near horizon geometry of the extremal rotating black hole in arbitrary dimension exhibits $SO(2,1)$ symmetry. Because Killing vectors are linked to first integrals of the geodesic equations,
the model of a massive relativistic particle propagating on such a background is
automatically conformal invariant. In this framework the generators of the conformal algebra schematically look as follows
\be\label{triple}
H=r \left( \sqrt{{(r p_r)}^2 + L(\mu,p_\mu,p_\phi)} -f(p_\phi) \right), \quad D_0=r p_r,\quad
K_0=\frac{1}{r} \left( \sqrt{{(r p_r)}^2 + L(\mu,p_\mu,p_\phi)}
+f(p_\phi) \right),
\ee
where the function $L(\mu,p_\mu,p_\phi)$ is at most quadratic in the momenta $p_{\mu_i}$, $p_{\phi_i}$ canonically conjugate to the angular variables $\mu_i$, $\phi_i$, while $f(p_\phi)$ is linear in the momenta. Their explicit form depends on the details of a black hole under consideration \cite{g1}-\cite{gn}. A comparison with (\ref{i}) gives\footnote{It is worth mentioning that for the Kerr black hole $L(\mu,p_\mu,p_\phi)$ can be linked to the near horizon Killing tensor of the second rank \cite{g2,go}.}
\be
\mathcal{I}=L(\mu,p_\mu,p_\phi)-f(p_\phi)^2.
\ee
However, with respect to the Poisson bracket the new
radial variables $(R, p_R)$ do not commute with $p_a=(p_\mu, p_{\phi}), \varphi^a=(\mu,\phi)$.
In order to split them, we perform a canonical transformation
$(r,p_r, \varphi^a, p_{a})\to
(R,p_R, {\widetilde\varphi^a}, {\widetilde p_{a}})$, which is defined by (\ref{Rtr})
and by the
following transformation of the remaining variables (for related earlier studies see \cite{bgik,g1,gn})
\be
\
{\widetilde\varphi^a}=\varphi^a+\frac{\partial U}{\partial p_a},\qquad
{\widetilde p_a}=p_a-\frac{\partial U}{\partial \varphi^a},\qquad U(rp_r,p_a,\varphi^a)\equiv
\frac12\int_{x=rp_r}dx\log\left(\sqrt{x^2/4 +L(p_a,\varphi^a)}
+f( p_a)\right).
\label{tphi}\ee
As a result, $(R,p_R)$ and $({\widetilde\varphi}^a, {\widetilde p}_a)$ constitute canonical pairs.
In contrast with the canonical transformation suggested in \cite{gn}, the one above does not appeal to a formulation in terms of action-angle
variables.

Thus,  by applying a proper canonical transformation one can bring the  model of
a massive relativistic particle moving near the horizon
of an extremal black hole to the conventional conformal mechanics form.
Important information about the system, which was originally defined in $d$ dimensions, is thus imprinted in the $(d-2)$--dimensional spherical mechanics, which derives from it.

For the extremal black hole with equal rotation parameters the Hamiltonian of a spherical mechanics was derived in \cite{gal}. In the case of $d=2n+1$ dimensions one finds
\bea\label{Hodd}
&&
\mathcal{I}=\sum_{i,j=1}^{n-1}(\delta_{ij}-\mu_i \mu_j) p_{\mu_i} p_{\mu_j}+
\sum_{i=1}^{n} \frac{p_{\phi_i}^2}{\mu_i^2} -\frac{(n+1)}{n} {\left(\sum_{i=1}^{n} p_{\phi_i} \right)}^2,
\eea
where $(\mu_i,p_{\mu_i})$, $i=1,\dots,n-1$ and $(\phi_j,p_{\phi_j})$, $j=1,\dots,n$ form canonical pairs obeying the conventional Poisson brackets $\{ \mu_i,p_{\mu_j}\}=\delta_{ij}$, $\{\phi_i,p_{\phi_j} \}=\delta_{ij}$ and $\mu_n^2$ entering the second sum in (\ref{Hodd}) is found from the unit sphere equation  $\sum_{i=1}^{n} \mu_i^2=1$. For $d=2n$ the Hamiltonian, which governs the corresponding spherical mechanics, reads
\bea\label{Heven}
&&
\mathcal{I}=\sum_{i,j=1}^{n-1}((2n-3) \rho_0^2 \delta_{ij}-\mu_i \mu_j) p_{\mu_i} p_{\mu_j}+
\sum_{i,j=1}^{n-1}\left(\frac{(2n-3)\rho_0^2}{\mu_i^2} \delta_{ij}-\frac{{(2n-3)}^2 \rho_0^2}{2(n-1)}-\frac{2}{n-1}
\right) p_{\phi_i} p_{\phi_j}+m^2 \rho_0^2,
\nonumber\\[2pt]
&&
\rho_0^2=\frac{2(n-1)}{2n-3}-\sum_{i=1}^{n-1} \mu_i^2,
\eea
where $(\mu_i,p_{\mu_i})$ and $(\phi_j,p_{\phi_j})$, $j=1,\dots,n-1$ form canonical pairs  and $m^2$ is a coupling constant. Note that, as compared to the previous case, the number of the azimuthal coordinates
is decreased by one.

Because the azimuthal angular variables $\phi_i$ are cyclic, it is natural to consider a reduction in which they are discarded. This is achieved by
setting in (\ref{Hodd}) and (\ref{Heven}) the momenta canonically conjugate to $\phi_i$ to be coupling constants
\be\label{red}
p_{\phi_i} \quad \rightarrow \quad g_i.
\ee
Note that, after such a reduction, both (\ref{Hodd}) and (\ref{Heven}) yield dynamical systems, which contain $(n-1)$ configuration space degrees of freedom.
The rest of this paper is devoted to a systematic study of the reduced models.

\section{Spherical mechanics related with black hole in $d=2n+1$}

For the spherical mechanics (\ref{Hodd}) associated with the extremal rotating black hole in $d=2n+1$ dimensions the reduction (\ref{red}) yields\footnote{We denote the reduced Hamiltonian by the same letter $\mathcal{I}$. This does not cause confusion, because, from now on, we abandon the parent formulations (\ref{Hodd}) and (\ref{Heven}).}
\bea\label{Hodd1}
&&
\mathcal{I}=\sum_{i,j=1}^{n-1}(\delta_{ij}-\mu_i \mu_j) p_{\mu_i} p_{\mu_j}+
\sum_{i=1}^{n}\frac{g_i^2}{\mu_i^2}, \qquad \mu_n^2=1-\sum_{i=1}^{n-1} \mu_i^2.
\eea
Since the first term in (\ref{Hodd1}) involves the inverse metric on an $(n-1)$--dimensional sphere,
the model can be interpreted as a particle moving on $\mathcal{S}^{n-1}$ in the external field.

The analysis of integrability of (\ref{Hodd1}) is facilitated in spherical coordinates. Introducing one angle at a time
\be
\mu_n=\cos\theta_{n-1},\quad \mu_i=x_i \sin\theta_{n-1}, \qquad \sum_{i=1}^{n-1} x^2_i=1
\label{transf}
\ee
and computing the metric induced on the sphere $\sum_{a=1}^n d \mu_a^2$ and its inverse, one can bring (\ref{Hodd1}) to the form
\bea\label{Hodd2}
&&
\mathcal{I}=p_{\theta_{n-1}}^2+\frac{g_n^2}{\cos^2{\theta_{n-1}}}+\frac{1}{\sin^2{\theta_{n-1}}} \left(
\sum_{i,j=1}^{n-2}(\delta_{ij}-x_i x_j) p_i p_j+
\sum_{i=1}^{n-1}\frac{g_i^2}{x_i^2}\right),
\\[2pt]
&&
x_{n-1}^2=1-\sum_{i=1}^{n-2} x^2_i,
\nonumber
\eea
where $p_i$ are momenta canonically conjugate to $x_i$, $i=1,\dots,n-2$. Thus, the canonical pair $(\theta_{n-1},p_{\theta_{n-1}})$ is separated, while the expression in braces gives the first integral of the Hamiltonian (\ref{Hodd2}). Because its structure is analogous to (\ref{Hodd1}), one can proceed along the same lines
\be
x_{n-1}=\cos\theta_{n-2},\quad x_a=y_a \sin\theta_{n-2}, \qquad \sum_{a=1}^{n-2} y^2_a=1
\label{transf}
\ee
until one achieves a complete separation of the variables. The resulting Hamiltonian is a kind of matryoshka doll
\bea\label{Hodd3}
&&
\mathcal{I}=F_{n-1},
\eea
where $F_{n-1}$ is derived from the recurrence relation
\be\label{F}
F_i=p_{\theta_i}^2+\frac{g_{i+1}^2}{\cos^2{\theta_{i}}}+\frac{F_{i-1}}{\sin^2{\theta_{i}}},
\ee
with $i=1,\dots,n-1$ and $F_0=g_1^2$. The functionally independent integrals of motion in involution $F_i$ ensure the integrability of (\ref{Hodd1}). To avoid confusion, let us stress that, given $n$, the Hamiltonian (\ref{Hodd3}) describes a system with $(n-1)$ configuration space degrees of freedom.
Note that in a different context this model has been discussed in \cite{vsp}. Worth mentioning also is that, if a system with the Hamiltonian $F_{i-1}$ has some integrals of motion, these automatically are the integrals of motion of a larger system governed by the Hamiltonian $F_i$. For $n=2$ Eq. (\ref{Hodd2}) reproduces the celebrated P\"oschl--Teller model \cite{pt}.

Although the integrability of (\ref{Hodd1}) is obvious in spherical coordinates, the fact that the model is maximally superintegrable is less evident. In order to prove it, we resort to the parent formulation (\ref{Hodd}) and analyze how the reduction (\ref{red})
affects the symmetries (\ref{vf}) \footnote{A realization of $U(n)$ in (\ref{Hodd}) is derived from Eq. (\ref{vf}) by the standard substitution
$\frac{\partial}{\partial \mu_i} \rightarrow p_{\mu_i}$, $\frac{\partial}{\partial \phi_i} \rightarrow p_{\phi_i}$, which links the Killing vectors to the first integrals of the Hamiltonian mechanics. The Hamiltonian (\ref{Hodd}) proves to be a combination of the first two Casimir elements
$\xi_{ij}^2+\rho_{ij}^2$ and ${(\rho_{ii})}^2$.}. First of all, we notice that $\rho_{ii}$ (no summation over repeated indices) generates rotation in the $i$--th plane. Within the canonical framework it is represented by $\rho_{ii}=2 p_{\phi_i}$. Then the very nature of the reduction mechanism (\ref{red}) suggests that those generators in (\ref{vf}), which Poisson commute with $\rho_{ii}$, will be symmetries of the reduced Hamiltonian (\ref{Hodd1}). Because (\ref{Hodd}) was constructed from the Casimir elements of $u(n)$, it is straightforward to verify that the combinations (no summation over repeated indices)
\be\label{I}
I_{ij}=\xi_{ij}^2+\rho_{ij}^2
\ee
with $i<j$  generate the desired symmetries.

Before we proceed to treat the general case, it proves instructive to illustrate the construction by the examples of $n=3$ and $n=4$, which correspond to  seven--dimensional and nine--dimensional black hole configurations. For $n=3$ the Hamiltonian reads\footnote{Here and in what follows the subscript attached to the Hamiltonian refers to the number of configuration space degrees of freedom in the model.}
\bea\label{Hodd4}
&&
{\mathcal{I}}_2=p_{\theta_{2}}^2+\frac{g_3^2}{\cos^2{\theta_{2}}}+\frac{1}{\sin^2{\theta_{2}}} \left(p_{\theta_1}^2+\frac{g_1^2}{\sin^2{\theta_1}}+\frac{g_2^2}{\cos^2{\theta_1}}
\right).
\eea
In order to construct the integrals of motion, one makes use of (\ref{planes}) and (\ref{vf})
\bea
&&
\xi_{12}=-p_{\theta_1} \cos{\phi_{12}} +\left(p_{\phi_1} \cot{\theta_1} +p_{\phi_2} \tan{\theta_1}\right) \sin{\phi_{12}},
\nonumber\\[4pt]
&&
\xi_{13}=-\left( p_{\theta_1} \cos{\theta_1} \cot{\theta_2} +p_{\theta_2}  \sin{\theta_1}\right) \cos{\phi_{13}} +
\left(p_{\phi_1} \frac{\cot{\theta_2}}{\sin{\theta_1}}+p_{\phi_3} \sin{\theta_1} \tan{\theta_2}\right) \sin{\phi_{13}},
\nonumber\\[4pt]
&&
\xi_{23}=\left( p_{\theta_1}  \sin{\theta_1} \cot{\theta_2} -p_{\theta_2} \cos{\theta_1}\right) \cos{\phi_{23}}+
\left(p_{\phi_2} \frac{\cot{\theta_2}}{\cos{\theta_1}}+p_{\phi_3} \cos{\theta_1} \tan{\theta_2}\right) \sin{\phi_{23}},
\nonumber\\[4pt]
&&
\rho_{12}=p_{\theta_1} \sin{\phi_{12}} +\left(p_{\phi_1} \cot{\theta_1} +p_{\phi_2} \tan{\theta_1}\right) \cos{\phi_{12}},
\nonumber\\[4pt]
&&
\rho_{13}=\left( p_{\theta_1} \cos{\theta_1} \cot{\theta_2} +p_{\theta_2}  \sin{\theta_1}\right) \sin{\phi_{13}} +
\left(p_{\phi_1} \frac{\cot{\theta_2}}{\sin{\theta_1}}+p_{\phi_3} \sin{\theta_1} \tan{\theta_2}\right) \cos{\phi_{13}},
\nonumber\\[4pt]
&&
\rho_{23}=-\left( p_{\theta_1}  \sin{\theta_1} \cot{\theta_2} -p_{\theta_2} \cos{\theta_1}\right) \sin{\phi_{23}}+
\left(p_{\phi_2} \frac{\cot{\theta_2}}{\cos{\theta_1}}+p_{\phi_3} \cos{\theta_1} \tan{\theta_2}\right) \cos{\phi_{23}},
\nonumber\\[4pt]
&&
\rho_{11}=2 p_{\phi_1}, \qquad \rho_{22}=2 p_{\phi_2}, \qquad \rho_{33}=2 p_{\phi_3},
\eea
where we abbreviated $\phi_{ij}=\phi_i-\phi_j$,  which after implementing the reduction (\ref{red}) yield
\bea\label{I1}
&&
{\tilde I}_{12}=p_{\theta_1}^2+\frac{g_1^2}{\sin^2{\theta_1}}+\frac{g_2^2}{\cos^2{\theta_1}},
\nonumber\\[4pt]
&&
{\tilde I}_{13}={(p_{\theta_1} \cos{\theta_1} \cot{\theta_2}+p_{\theta_2} \sin{\theta_1})}^2+{\left(g_1 \frac{\cot{\theta_2}}{\sin{\theta_1}}+g_3 \sin{\theta_1} \tan{\theta_2} \right)}^2,
\nonumber\\[4pt]
&&
{\tilde I}_{23}={(p_{\theta_1} \sin{\theta_1} \cot{\theta_2}-p_{\theta_2} \cos{\theta_1})}^2+{\left(g_2 \frac{\cot{\theta_2}}{\cos{\theta_1}}+g_3 \cos{\theta_1} \tan{\theta_2} \right)}^2.
\eea
It is straightforward to verify that the vectors $\partial_A {\tilde I}_{ij}$,  where $A=(\theta_1,\theta_2,p_{\theta_1},p_{\theta_2})$ are linearly independent and, hence, the first integrals are functionally independent. Because the Hamiltonian is constructed from ${\tilde I}_{ij}$ \footnote{Recall that the parent Hamiltonian (\ref{Hodd}) was constructed from the Casimir elements of $u(n)$. Up to a constant, the sum $\sum_{i<j=1}^n {\tilde I}_{ij}$ is what is left after the reduction.}
\bea
&&
{\mathcal{I}}_2={\tilde I}_{12}+{\tilde I}_{13}+{\tilde I}_{23}+g_3 (g_3-2 g_1 - 2 g_2),
\eea
one has three functionally independent integrals of motion for a system with two degrees of freedom and, hence, the model is maximally superintegrable. Note that the algebra formed by ${\tilde I}_{ij}$ is nonlinear.
It is convenient to treat the Hamiltonian  ${\mathcal{I}}_2$ (with the additive constant $g_3 (g_3-2 g_1 - 2 g_2)$ being discarded) and ${\tilde I}_{12}$ as the first integrals in involution, while ${\tilde I}_{23}$ is the additional first integral, which renders the model maximally superintegrable.

The case $n=4$ is treated likewise. From Eqs. (\ref{Hodd3}) and (\ref{F}) one derives the Hamiltonian
\bea
&&
{\mathcal{I}}_3=p_{\theta_3}^2+\frac{g_4^2}{\cos^2{\theta_3}}+\frac{1}{\sin^2{\theta_3}} \left[ p_{\theta_2}^2+\frac{g_3^2}{\cos^2{\theta_2}}+\frac{1}{\sin^2{\theta_2}} \left(p_{\theta_1}^2+\frac{g_2^2}{\cos^2{\theta_1}}+\frac{g_1^2}{\sin^2{\theta_1}} \right)\right],
\nonumber\\[2pt]
&&
\eea
while the first integrals prove to be exhausted by those in (\ref{I1}) and three more functions
\bea\label{I3}
&&
{\tilde I}_{14}={\left(p_{\theta_1} \frac{\cos{\theta_1} \cot{\theta_3}}{\sin{\theta_2}}+p_{\theta_2} \sin{\theta_1} \cos{\theta_2} \cot{\theta_3}+p_{\theta_3} \sin{\theta_1} \sin{\theta_2}\right)}^2+\left(g_1 \frac{\cot{\theta_3}}{\sin{\theta_1} \sin{\theta_2}}+
\right.
\nonumber\\[2pt]
&& \qquad \quad
{\left.
g_4 \sin{\theta_1} \sin{\theta_2}  \tan{\theta_3} \frac{}{} \right)}^2,
\nonumber\\[2pt]
&&
{\tilde I}_{24}={\left(p_{\theta_1} \frac{\sin{\theta_1} \cot{\theta_3}}{\sin{\theta_2}}-p_{\theta_2} \cos{\theta_1} \cos{\theta_2} \cot{\theta_3}-p_{\theta_3} \cos{\theta_1} \sin{\theta_2}\right)}^2+\left(g_2 \frac{\cot{\theta_3}}{\cos{\theta_1} \sin{\theta_2}}+
\right.
\nonumber\\[2pt]
&& \qquad \quad
{\left.
g_4 \cos{\theta_1} \sin{\theta_2}  \tan{\theta_3} \frac{}{} \right)}^2,
\nonumber\\[2pt]
&&
{\tilde I}_{34}={\left(p_{\theta_2} \sin{\theta_2} \cot{\theta_3} -p_{\theta_3} \cos{\theta_2} \frac{}{}\right)}^2+{\left(g_3 \frac{\cot{\theta_3}}{\cos{\theta_2}}+g_4 \cos{\theta_2} \tan{\theta_3} \right)}^2.
\eea
As in the preceding case, the Hamiltonian is a combination of ${\tilde I}_{ij}$
\be
{\mathcal{I}}_3=\sum_{i<j}^4{\tilde I}_{ij}+{(g_3-g_4)}^2-2 g_1 (g_3+g_4)-2 g_2 (g_3+g_4).
\ee
Because for a system with $n$ configuration space degrees of freedom the maximal number of functionally independent integrals of motion is $2n-1$, the set (\ref{I1}) and (\ref{I3}) is overcomplete and only five functions prove to be independent.

That for generic $n$ the model is maximally superintegrable can now be proved by induction. For $n=2$ the systems involves only one configuration space degree of freedom and the Hamiltonian is the only integral of motion. For $n=3$ we choose ${\mathcal{I}}_2$, ${\tilde I}_{12}$ and ${\tilde I}_{23}$ to be the functionally independent first integrals.
When passing from $n=3$ to $n=4$, the integrals of motion of the former model are automatically the integrals of motion of the latter. To complete the set, we choose ${\mathcal{I}}_3$ and ${\tilde I}_{34}$. Obviously, this process can be continued to any order. Given a superintegrable system with the Hamiltonian ${\mathcal{I}}_{n-1}$, $n-1$ configuration space degrees of freedom and $2(n-1)-1$ functionally independent integrals of motion, one introduces one more configuration space degree of freedom and two new integrals of motion ${\mathcal{I}}_{n}$ and
\be
{\tilde I}_{n-1, n}={\left(p_{\theta_{n-2}} \sin{\theta_{n-2}} \cot{\theta_{n-1}} -p_{\theta_{n-1}} \cos{\theta_{n-2}} \frac{}{}\right)}^2+{\left(g_{n-1} \frac{\cot{\theta_{n-1}}}{\cos{\theta_{n-2}}}+g_n \cos{\theta_{n-2}} \tan{\theta_{n-1}} \right)}^2,
\ee
which all together describe a system with $n$ configuration space degrees of freedom and $2n-1$ functionally independent integrals of motion.

Let us construct the action--angle variables for the system. Following the standard procedure \cite{arnold}, one introduces the generating function
\be\label{Sodd}
S^{odd}(F_i,|g_i|,\theta_i)=\sum_{i=1}^{n-1}\int p_{\theta_i}(F_1,\dots,F_{n-1}, \theta_i) \mathrm{d}\theta_i,
\ee
where $p_{\theta_i}(F_1,\dots,F_{n-1}, \theta_i)$  are to be expressed from  (\ref{F}).
For the action variables one has\footnote{For technical details on evaluating the integrals see e.g. Ref. \cite{hlnsy}.}
\be\label{Iodd}
I_i=\frac{1}{2 \pi} \oint \mathrm{d}\theta_i \left[ \sqrt{F_i-\frac{F_{i-1}}{\sin^2\theta_i}-\frac{g_{i+1}^2}{\cos^2\theta_i}}\right] =\frac12(\sqrt{F_i}-\sqrt{F_{i-1}}-|g_{i+1}|),
\ee
which can be inverted to yield
\be
F_i=\left(2 \sum_{k=1}^i I_k + \sum_{k=1}^{i+1}|g_{k}|\right)^2.
\label{Fodd}\ee
The angle variables are defined by
\be
\Phi^{odd}_i=\frac{\partial S^{odd}}{\partial I_i}=\sum_{k=i}^{n-1} \arcsin X_k+2 \sum_{k=i+1}^{n-1} \arctan Y_k,
\label{Phiodd}\ee
where we abbriviated
\be
\begin{array}{ccl}
&&X_k  =  \frac{\left(F_k+F_{k-1}-g_{k+1}^2\right)-2 F_k \sin^2 \theta_k}{\sqrt{\left(-F_k+F_{k-1}-
g_{k+1}^2\right)^2-4 F_k g_{k+1}^2}}\\
&&Y_k
 =2\frac{
{\left(F_k+F_{k-1}-{g}_{k+1}^2\right)}
\sqrt{{F_k \sin^2 \theta_k \cos^2 \theta_k - F_{k-1} \cos^2 \theta_k
 -{g}_{k+1}^2\sin^2 \theta_k}}-{ \sin^2 \theta_k}\sqrt{{F_k\left(F_k+F_{k-1}-{g}_{k+1}^2\right)^2- F^2_kF_{k-1}}}}
 {\sqrt{F_{k-1}}\left(F_k+F_{k-1}-{g}_{k+1}^2)-2 F_k \sin^2 \theta_k\right)}

\end{array}
\label{XY}\ee
Being rewritten in the action--angle variables, the Hamiltonian reads
\be
{\mathcal{I}}= \left(2 \sum_{k=1}^{n-1} I_k + \sum_{k=1}^n|{g}_{k}|\right)^2,
\ee
which coincides with the Hamiltonian of a free particle on an $(n-1)$--dimensional sphere up to the shift of the action variables \cite{hlnsy}. Thus, the only difference with that case is the shift in the range of
$\sum_k I_k$ from $[0,\infty )$ to $[\sum_{k=1}^n|\hat{g}_{k}|, \infty ) $. Thus, the system possesses $SO(n+1)$ symmetry and is, obviously, maximally superintegrable.

Let us discuss how hidden constants of the motion can be revealed within the action--angle formulation. Evolution of the angle variables is governed by the equation (see, e.g., Refs. \cite{hlnsy,gonera})
\be
\frac{d \Phi^{odd}_i}{d t}=2\left(2 \sum_{k=1}^{n-1} I_k + \sum_{k=1}^n|{g}_{k}|\right).
\ee
The expressions $\cos(\Phi^{odd}_i-\Phi^{odd}_j+ {\rm const})$ define constants of the motion for any $i,j=1,\dots,n-1$ and only
$n-2$ of these are  functionally independent
\be
G_i=\cos\left(\Phi^{odd}_i-\Phi^{odd}_{i+1}\right)
=\frac{\sqrt{1-X_i^2}(1-Y_{i+1}^2)-2 X_i Y_{i+1}}{1+Y_{i+1}^2},
\label{hiddenAA}\ee
where $i=1,\dots,n-2$. Because the $(n-1)$--dimensional system has $(2n-3)$ functionally independent constants of the motion, it is maximally superintegrable.
The fact that the Hamiltonian is expressed via the action variables in terms of elementary functions implies also that the system is exactly solvable.

\section{Spherical mechanics related with black hole in $d=2n$}

For the spherical mechanics (\ref{Heven}) associated with the extremal rotating black hole in $d=2n$ dimensions the reduction (\ref{red}) yields
\bea\label{Final}
&&
\mathcal{I}=\sum_{i,j=1}^{n-1}((2n-3) \rho_0^2 \delta_{ij}-\mu_i \mu_j) p_{\mu_i} p_{\mu_j}+\sum_{i=1}^{n-1} \frac{(2n-3) \rho_0^2 g_i^2 }{\mu_i^2}+\nu \sum_{i=1}^{n-1} \mu_i^2,
\eea
where $\nu$ and $g_i$ are coupling constants and $\rho_0^2$ is given in (\ref{Heven}).

Like above, the proof of superintegrability of (\ref{Final}) is facilitated by introducing the spherical coordinates
\be
\mu_i=x_i \sin{\theta_{n-1}}, \qquad \sum_{i=1}^{n-1} x_i^2=1 \quad \Rightarrow \quad \sum_{i=1}^{n-1} \mu_i^2=\sin^2{\theta_{n-1}}.
\ee
In order to transform the kinetic term in (\ref{Final}), one inverts the metric then computes the line element in spherical coordinates and then inverts it again.
This yields
\be\label{HH}
\mathcal{I}=2(n-1) p^2_{\theta_{n-1}}+\nu \sin^2{\theta_{n-1}}+\left(\frac{2(n-1)}{\sin^2{\theta_{n-1}}}-2n+3 \right)
\left(
\sum_{i,j=1}^{n-2}(\delta_{ij}-x_i x_j) p_i p_j+
\sum_{i=1}^{n-1}\frac{g_i^2}{x_i^2}\right),
\ee
where $p_i$ are momenta canonically conjugate to $x_i$, $i=1,\dots,n-2$. Beautifully enough, the rightmost factor in (\ref{HH}) is the Hamiltonian of a particle on ${\mathcal{S}}^{n-2}$, which was studied in detail in the preceding Section. This sector provides $2(n-2)-1$ functionally independent integrals of motion, which correlates with the $U(n-1)$ symmetry of the parent formulation (\ref{Heven}). Because (\ref{HH}) involves one more canonical pair $(\theta_{n-1},p_{\theta_{n-1}})$ and only one extra integral of motion (the Hamiltonian (\ref{HH}) itself), the full theory lacks for only one integral of motion to be maximally superintegrable.

Let us construct action--angle variables for the system. In order to simplify the bulky formulae below, from now on we change the notation
$n \to n+1$, which corresponds to a black hole in $d=2(n+1)$ dimensions. To avoid confusion, the corresponding Hamiltonian will be denoted by ${\cal I}_{0}$
\be
{\cal I}_0=2 n p^2_{\theta_{n}}+\nu\sin^2\theta_{n} + \left(\frac{2 n}{\sin^2 \theta_{n}}-2n+1\right)F_{n-1},
\ee
with
$F_{n-1}$ given in (\ref{F}).
One starts with the generating function
\be
S^{even}=\sum_{i=1}^{n}\int p_{\theta_i}({\cal I}_0, F_1,\dots,F_{n-1}, \theta_i) \mathrm{d}\theta_i=
\int p_{\theta_n}({\cal I}_0, F_{n-1}, \theta_n)d\theta_n +S^{odd},
\label{Seven}\ee
where $S^{odd}$ has the structure similar to (\ref{Sodd}),
and the expression for $p_{\theta_n}$  is derived from the Hamiltonian ${\cal I}_{0}$.
The action variables $I_1,\ldots I_{n-1}$ coincide with those in the odd--dimensional case, while
for $I_{n}$ one gets
\be
I_{n}=\sqrt{\frac{-a^- \nu} {8n}}a^+ {\cal F}_1\left(\frac12,1,-\frac12,2,a^+,\frac{a^+}{a^-}\right),
\label{55}\ee
where ${\cal F}_1$ is Appell's first hypergeometric function (see e.g. \cite{dwight}) and
\be
a^\pm=1-\frac{{\cal I}_0}{2 \nu}-\frac{2n-1}{2 \nu}F_{n-1} \pm \sqrt{\left(1-\frac{{\cal I}_0}{2 \nu}-\frac{2n-1}{2 \nu}F_{n-1}\right)^2+\frac{{\cal I}_0}{\nu}-\frac{F_{n-1}}{\nu}-1}.
\label{56}\ee
Inverting this expressions, we would get the
Hamiltonian written in terms of the action variables. Unfortunately, this cannot be done in elementary functions.
While the system under consideration is integrable, it fails to be exactly solvable.

The angle variable conjugated to $I_n$  reads
\be
\Phi^{even}_{n}=\frac{\partial {\cal I}_0}{\partial I_{n}}\frac1{\sqrt{8 \nu n a^+}} {\cal F}\left(\arcsin\sqrt{\frac{a^+}{a^+-\cos^2 \theta_{n}}}, 1-\frac{a^-}{a^+}\right),
\ee
while other $(n-1)$ angle variables are defined by the expressions
\be
\Phi^{even}_i=\Phi^{odd}_i- A \Pi\left(1-\frac1{a^+},\arcsin\sqrt{\frac{a^+}{a^+-\cos^2 \theta_{n}}}, 1-\frac{a^-}{a^+}\right)+ B  {\cal F}\left(\arcsin\sqrt{\frac{a^+}{a^+-\cos^2 \theta_{n}}}, 1-\frac{a^-}{a^+}\right),
\ee
where $\Phi^{odd}_i$ were defined in the preceding section, ${\cal F}(\phi | m)$ is the elliptic integral of the first kind,
$\Pi(n; \phi | m)$ is the elliptic integral of the third kind, and we abbriviated
\be A=\sqrt{\frac{8 n F_{n-1}}{\nu}}\frac{1}{\sqrt{a^+}(a^+-1)},\quad
B=A+\frac{\sqrt{2 F_{n-1}}}{\sqrt{n \nu a^+}}\left(\frac{\partial {\cal I}_0}{\partial F_{n-1}}+2 n-1\right).
\ee
It follows  from (\ref{55}) and (\ref{56}), that the  ratio of the effective frequencies ${\omega_1}=\partial{\cal I}/\partial{I_{n}}$ and ${\omega_2}=\partial{\cal I}/\partial{F_{n-1}}$ is not a rational number. Furthermore, it is a function of the action variables.
Hence, although $(\omega_2\Phi_n - \omega_1\Phi_i)$ commute with the Hamiltonian ${\cal I}_0$, they are not periodic. As a result, using these functions one cannot define additional globally defined constants of the motion (for a related discussions see \cite{gonera,hlnsy}).
All hidden symmetries of the model are thus contained in (\ref{hiddenAA}). Because the
$n$-dimensional system has $n+(n-2)=2n-2$ constants of the motion, it lacks for only one first integral to be maximally superintegrable system.

\section{Concluding remarks}

To summarize, in this work we provided a systematic account of integrability of spherical mechanics models associated with
the near horizon extremal Myers--Perry black hole in arbitrary dimension for the special case that all rotation parameters are equal. The integrability was established both in the original coordinates and in action--angle variables.
It was demonstrated that the spherical mechanics associated with the black hole in $d=2n+1$ dimensions is maximally superintegrable, while its counterpart related to the black hole in $d=2n$
lacks for only one constant of the motion to be maximally superintegrable.

Our analysis implies that the parent formulations (\ref{Hodd}) and (\ref{Heven}) are superintegrable as well.
Indeed, bearing in mind the reduction formula (\ref{red}), one can consider the generating functions for the parent formulation $S^{odd,even} (F_i,\theta_i,p_{\phi_i})+\sum_i p_{\phi_i}\phi_i$,
where the first term is defined either by (\ref{Sodd}) or by (\ref{Seven}), with $g_i$ being replaced by $p_{\phi_i}$. From here its follows that the action variables for the parent systems are given by $I_i$ and $p_{\phi_i}$.
The angle variables corresponding to $I_i$ are the same as in the reduced models, while those corresponding to $p_{\phi_i}$ are $\phi_i+\partial{S^{odd,even} }/\partial p_{\phi_i}$.
Repeating the same arguments as for the reduced systems, one can verify that the parent formulation related to the black hole in $d=2n+1$ is maximally superintegrable, while that associated with the black hole in $d=2n$
lucks for one integral of motion to become maximally superintegrable.

\vspace{0.5cm}

\noindent
{\large Acknowledgements.} This work was supported by RF Federal Program "Kadry" under the contract 16.740.11.0469, MSE Program "Nauka" under the grant 1.604.2011, DFG grant LE 838/12-1 (A.G.), the Armenian State Committee of Science  grant 11-1c258 and by Volkswagenstiftung under the contract nr. 86 260 (A.N., A.S.).

\end{document}